\documentclass[twocolumn,showpacs,preprintnumbers,amsmath,amssymb]{revtex4}

\usepackage{amsmath}
\usepackage{amssymb}

\usepackage{graphicx}
\usepackage{dcolumn}
\usepackage{bm}

\begin{document}


\title{Ordering in the pyrochlore antiferromagnet due to Dzyaloshinsky-Moriya interactions.}

\author{Maged Elhajal}
\email{melhajal@mpi-halle.de}
\altaffiliation[\newline Present address~: ]{Max Planck Institut, Weinberg 2, 06120 Halle, Germany}
\author{Benjamin Canals}
\author{Raimon Sunyer i Borrell}
\altaffiliation[Present address~: ]{Unitat de Biof\'\i sica i Bioenginyeria, Facultat de Medicina, Universitat de Barcelona, Casanova 143, Barcelona 08036, Spain}
\author{Claudine Lacroix}
\affiliation{
Laboratoire Louis N\'eel, CNRS, 25 avenue des Martyrs, BP 166, 38042 Grenoble Cedex 9, France
}%

\date{\today}

\begin{abstract}
The Heisenberg nearest neighbour antiferromagnet on the pyrochlore (3D) lattice is highly frustrated and does not order at low temperature where spin-spin 
correlations remain short ranged. 
Dzyaloshinsky-Moriya interactions (DMI) may be present in pyrochlore compounds as is shown, and the consequences of such interactions on the magnetic properties are 
investigated through mean field approximation and monte carlo simulations. 
It is found that DMI (if present) tremendously change the low temperature behaviour of the system.
At a temperature of the order of the DMI a phase transition to a long range ordered state takes place. 
The ordered magnetic structures are explicited for the different possible DMI which are introduced on the basis of symmetry arguments. 
The relevance of such a scenario for pyrochlore compounds in which an ordered magnetic structure is observed experimentally is dicussed. 
\end{abstract}

\maketitle

\section{Introduction}

\begin{figure}
\includegraphics[width=75mm]{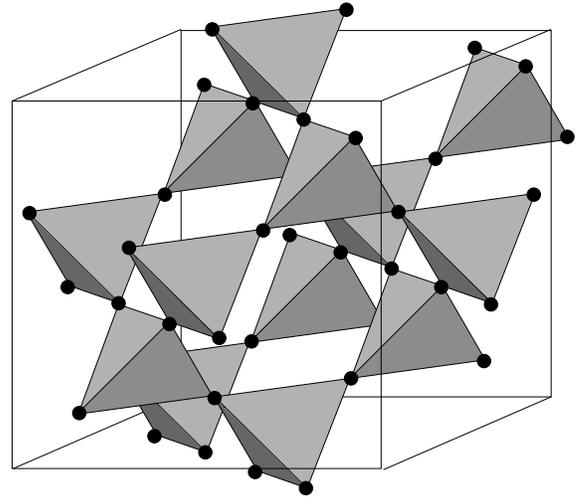}
\caption{\label{fig1}The pyrochlore lattice. The interplay of the frustration of the tetrahedral unit cell with the weak connectivity of the (corner sharing) tetrahedra provides peculiar magnetic properties 
to the pyrochlore antiferromagnet.}
\end{figure}

Frustration in magnetic systems can lead to unconventional magnetic ground states and peculiar low temperature behaviours \cite{Ramirez}. 
For example, in two-dimensional and three-dimensional highly frustrated systems, the conventional N\'eel ground state may be destabilized if frustration is 
associated to a weak connectivity of the lattice. This is the case for example in the kagom\'e and the pyrochlore \cite{Harris} lattices (see Fig. \ref{fig1}). 
In these lattices, in the presence of nearest neighbour antiferromagnetic interactions, the ground state is a disordered spin liquid state characterized by the 
abscence of magnetic long range order. 

The classical Heisenberg model on these two lattices has a macroscopic degeneracy of the ground state \cite{ReimersBerlinsky} which prevents any N\'eel-like ordering at $T=0$.
At small but finite temperature, there is a tendancy to coplanar order in the kagom\'e lattice \cite{Chalker} but no order of any kind is observed in the pyrochlore lattice 
\cite{Reimers, MoessnerPRL}.
For the pyrochlore lattice, in the extreme quantum limit of spins $S=\frac{1}{2}$, there is clear evidence for the absence of long range spin-spin correlations 
(and thus the abscence of N\'eel-like ordering) \cite{CanalsPRL}.
However, the nature of the disordered state (resonating valence bond spin liquid, valence bond crystal,\ldots) is still unclear \cite{HarrisBerlinsky, Hfm2000, gtdc1, gtdc2, Berg}.

On the experimental side, the pyrochlore antiferromagnets either do not order at low temperature, or have a critical temperature much smaller than the Curie-Weiss temperature $\theta_{CW}$, which is a general 
trend for highly frustrated compounds \cite{Ramirez}.

The reason why a number of pyrochlore compounds order at low temperature is that the spin liquid state is very sensitive to any additional term in the hamiltonian 
such as on-site anisotropies \cite{Champion}, dipolar interactions \cite{Palmer}, biquadratic interactions, next nearest neighbour interactions\ldots
Indeed, nearest neighbour antiferromagnetic interactions leave a high degeneracy of the ground state, or at least a high density of states at low energy. 
If the additional interactions are able to lift this degeneracy, they will be primarily responsible for the low temperature behaviour of the system. 

In this article, we consider the effect of small Dzyaloshinsky-Moriya interactions (DMI) in addition to nearest neighbour antiferromagnetic interactions. 
The same study was done on the kagom\'e lattice \cite{DM_Kagome} and (as it will be clear from the next sections) similar consequences of DMI were found, although the 
magnetic structure is different in the two cases.
In this paper, spins are treated as classical variables whereas a recent work \cite{Valeri} considers the extreme quantum limit of spins S=$\frac{1}{2}$, using another method. 
Unlike the present approach, quantum fluctuations are taken into account, but other approximations are made \cite{Valeri}.
Both approaches give to some extent similar (although not identical) results, which makes the overall picture of DMI induced ordering more reliable since the two methods start from different limits and use 
different approximation schemes.

In section \ref{IntroDMI}, the microscopic origin of DMI is presented and it is shown how to take them into account for pyrochlore systems, in accordance with the symmetry of the lattice.
Section \ref{MagnStruc} explicits the two magnetic structures obtained for the two possible DMI we have considered, and the possible link with experimental results is 
done. 

\section{\label{IntroDMI}DMI and Moriya's rules for the pyrochlore lattice} %

Taking into account Coulomb repulsion and Pauli's principle, Anderson \cite{Anderson} has explicited a microscopic mecanism which leads to isotropic 
super-exchange interactions ($J\mathbf{S}_{1}.\mathbf{S}_{2}$) in insulators. 
The main idea is that the energy is lowered if the spins of the electrons on neighbouring sites are in oposit directions because they can then minimize their kinetic 
energy by delocalizing on the nearest atoms, which is forbidden by Pauli principle if the spins are parallel. 
This fact is translated in an effective magnetic hamiltonian~: $J\mathbf{S}_{1}.\mathbf{S}_{2}$. 

Taking into account a weak spin-orbit coupling ($\lambda\mathbf{L}.\mathbf{S}$) and expanding in powers of $\lambda$, Moriya \cite{MoriyaPRL, MoriyaPR} showed 
that the effective magnetic hamiltonian between two spins $\mathbf{S}_{1}$ and $\mathbf{S}_{2}$ has the following expression~: 

\begin{equation}
\label{22h26}
H_{\text{mag}}=J\mathbf{S}_{1}.\mathbf{S}_2+
\mathbf{D}_{12}.\left(\mathbf{S}_{1}\times\mathbf{S}_2\right)+
\mathbf{S}_{1}.\overline{\overline{\Gamma}}.\mathbf{S}_{2}+\ldots
\end{equation}

\noindent with $J \propto \lambda^{0}$ whereas $D\propto\lambda$ and $\overline{\overline{\Gamma}}$ which is a symmetrical tensor is $\propto \lambda^{2}$. 
These are the leading orders in $\lambda$ and the second term of (\ref{22h26}) is the Dzyaloshinsky-Moriya interaction (DMI). 
Despite their higher order in $\lambda$, the third term and higher order terms should \emph{not always} be neglected compared to the DMI for two reasons. 
First, taking all the orders into account can restore the rotationnal symmetry of the Hamiltonian as was shown in \cite{Shekhtman, Kaplan}. 
This requires however some particular symmetry properties of the lattices \cite{Shekhtman} which do not hold for the pyrochlore lattice. 
The second case where higher order terms should not be neglected is when the magnetic structure induced by the isotropic interaction (J) is essentially collinear 
(ferromagnetic or antiferromagnetic) since the cross products $\mathbf{S}_i\times\mathbf{S}_j$ will then be of the order of $\sim\lambda$, and the DMI term in (\ref{22h26}) will be of 
order $\sim\lambda^2$, the same order as the next term ($\overline{\overline{\Gamma}}$ in eq. (\ref{22h26})).
However, collinear structures do not play any particular role in the pyrochlore lattice \cite{MoessnerPRL} and we shall only consider the first two terms of 
(\ref{22h26}) in this paper. 

Moriya's microscopic derivation of the DMI is only valid for insulators but other possible microscopic mecanism relevant for other materials were explicited, for 
instance in systems with RKKY interactions \cite{Fert, Levy}. 

Whatever the microscopic origin of the DMI is, there are always symmetry constraints on the possible \textbf{D} vectors which may appear in the 
hamiltonian. 
Indeed, the hamiltonian must be invariant under the symmetry operations of the crystal and this will restrict the possible \textbf{D} vectors to 
those for which the expression (\ref{22h26}) is invariant (under these symmetry operations). 
This way of constraining the \textbf{D} vectors has been given the name of Moriya's rules \cite{MoriyaPR}. 
We will make use of these rules to determine the possible DMI in the pyrochlore lattice. 

\begin{figure}
\includegraphics[width=85mm]{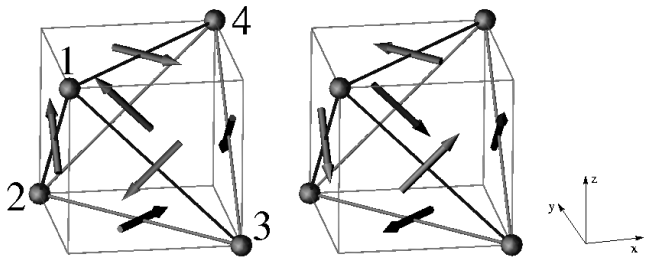}
\caption{\label{fig2}\textbf{D} vectors for the DMI in the pyrochlore lattice. 
The convention is taken to fix the order for the cross products (always $\mathbf{D}.\mathbf{S}_i\times\mathbf{S}_j$ with $j>i$).
The two possible DMI are those obtained by varying the direction of the \textbf{D} vectors ($\mathbf{D}\to-\mathbf{D}$).
The case with the \textbf{D} as represented on the left is refered to as the ``direct'' case and the other case (on the right) is the ``indirect'' case.
Once the DMI between two spins is fully specified, the others DMI are also fixed and obtained by applying the different $\frac{2\pi}{3}$ 
rotations around the cube's diagonals which leave the tetrahedron invariant.
The DMI in the rest of the lattice are also fixed and obtained by applying appropriate symmetry operations of the pyrochlore lattice (see text and figure \ref{fig3}).}
\end{figure}

One of Moriya's rules states that if the point between two magnetic sites is a center of inversion, then there can not exist a DMI. 
For that reason DMI are absent in crystals with a high symmetry, but this does not rule out DMI in the pyrochlore lattice which has no inversion center at the 
middle point between two sites. 
Carrying on the symmetry analysis one can in fact determine completely the direction of the \textbf{D} vectors as follows. 
Considering a single tetrahedron (the pyrochlore lattice is an assembly of corner sharing tetrahedra), the plane which contains two sites and the middle point of 
the opposite bond in the tetrahedron is a mirror plane (these planes are \{110\} planes in the cubic cell of figure \ref{fig1}). 
Applying Moriya's rules then implies that the \textbf{D} vector can only be perpendicular to this plane, or equivalently parallel to the opposite bond as shown on 
figure \ref{fig2}. 
There are so two possible DMI between two sites which correspond to the two directions for the \textbf{D} vector (and keeping the same 
order for the cross-product $\mathbf{S}_{i}\times\mathbf{S}_{j}$). 

The symmetry analysis (Moriya's rules) can not give more information about DMI.
Indeed, the \emph{sign} of \textbf{D} along the direction previously determined depends on the microscopic details of the particular compound \cite{MoriyaPR}, such as the number of electrons on the magnetic 
sites, the occupied orbitals, the crystal field...
At present, there is no general scheme to determine \textbf{D} quantatively, except in simple models which are unrealistic for the compounds we consider later on.
For this reason, we will consider the \emph{only two possible} DMI as determined by symmetry.

One can equivalently describe the two possible DMI with a fixed \textbf{D} vector but changing the order of the cross product since~: 
$\mathbf{D}.(\mathbf{S}_i\times\mathbf{S}_j)=-\mathbf{D}.(\mathbf{S}_j\times\mathbf{S}_i)$. 
This is done without any loss of generality since changing simultaneously the direction of \textbf{D} and the order of the cross product 
leaves the whole interaction unchanged. 
In the following the signs of the \textbf{D} vectors are changed while keeping always the same order for the cross products $\mathbf{S}_{i}\times\mathbf{S}_{j}$.
These two possibilities are refered to as ``direct'' and ``indirect'' DMI and are explicited in figure \ref{fig2}.

\begin{figure}
\includegraphics[height=4cm]{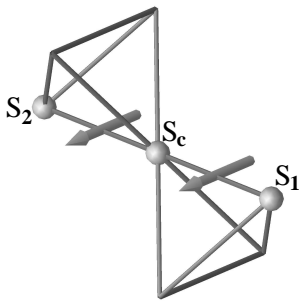}
\caption{\label{fig3}Symetry rules to deduce the \textbf{D} vectors from one tetrahedron to its neighbours. 
The point where the spin $\mathbf{S}_c$ is located is a center of inversion which leaves the whole lattice unchanged, and the hamiltonian must be invariant in this symetry operation.
$\mathbf{S}_2$ is the image of $\mathbf{S}_1$ and thus given the interaction $\mathbf{D}_{c1}.(\mathbf{S}_c\times \mathbf{S}_1)$, the interaction $\mathbf{D}_{c2}.(\mathbf{S}_c\times \mathbf{S}_2)$ is easily 
deduced.
The two $\mathbf{D}$ vectors are represented on the figure.}
\end{figure}

Once one particular DMI between two spins is fully specified, all the other DMI in the lattice are also fixed by symmetry. 
For instance the DMI within one tetrahedron can be deduced one from another thanks to the four three-fold rotation axes. 
The result is shown in figure \ref{fig2}. 
The DMI in the rest of the lattice are obtained by applying appropriate symmetries of the lattice as follows. 
Two neighbouring tetrahedra in the pyrochlore lattice are corner sharing and are transformed one into another by the inversion center located at the common site (site $\mathbf{S}_c$ on figure \ref{fig3}). 
The DMI hamiltonian must be invariant in this symmetry operation, and thus the DMI on one tetrahedron are easily obtained from the DMI on a neighbouring tetrahedron 
(see figure \ref{fig3}). 

\section{\label{MagnStruc}Magnetic properties} %

In this section, the consequences of DMI on the magnetic properties of a pyrochlore antiferromagnet are studied within mean field 
approximation and by classical monte carlo simulations. 
Monte carlo simulations indicate a phase transition to a long range ordered magnetic structure even for very weak DMI compared to 
antiferromagnetic isotropic exchange (J).
This is seen on the specific heat which seems to show a singularity when increasing system size (see figure \ref{fig4}) as well as on snapshots of the magnetic 
structure at 
very low temperature.
Thus, the low temperature structure is in deep contrast with the spin disordered state in the abscence of DMI \cite{CanalsPRL, 
MoessnerPRL}. 
We start in section \ref{direct} with the results in the case of direct DMI.
The case of indirect DMI is considered in section \ref{indirect}. 
The ordering of some pyrochlore compounds which could be due to DMI is then discussed. 
A slightly different geometry of the \textbf{D} together with some additional second nearest neighbour antiferromagnetic interactions is considered in relation with experimental results on mineral 
paramelaconite (Cu$_{4}$O$_{3}$) \cite{Pinsard}. 

\subsection{\label{direct}direct DMI}

We start with the case of direct DMI (see section \ref{IntroDMI} for a definition of ``direct''). 
We first show that DMI lead to a phase transition at low temperature, and then explicit the ordered magnetic structure. 

\subsubsection{Phase transition}

\begin{figure}
\includegraphics[width=85mm]{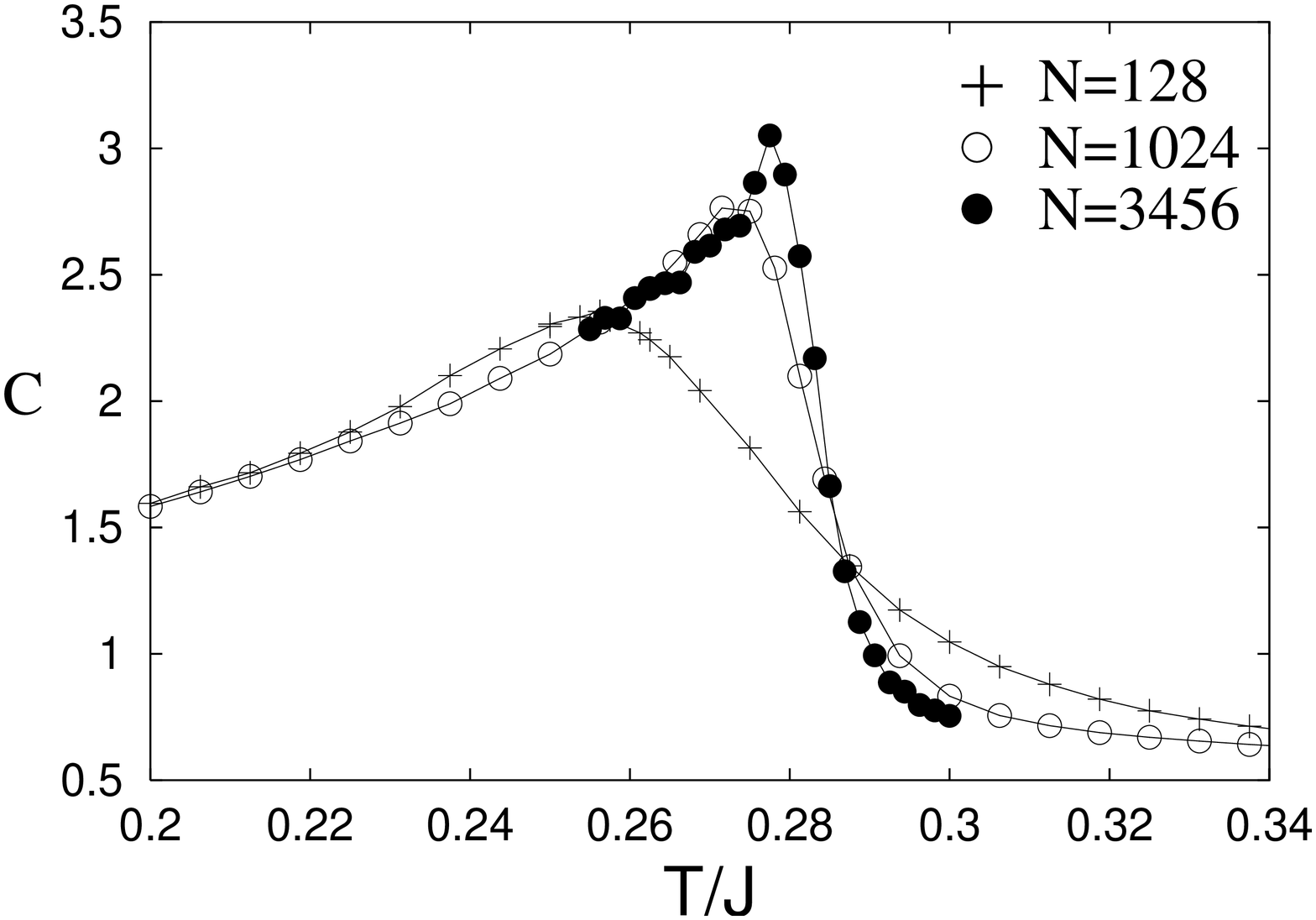}
\includegraphics[width=85mm]{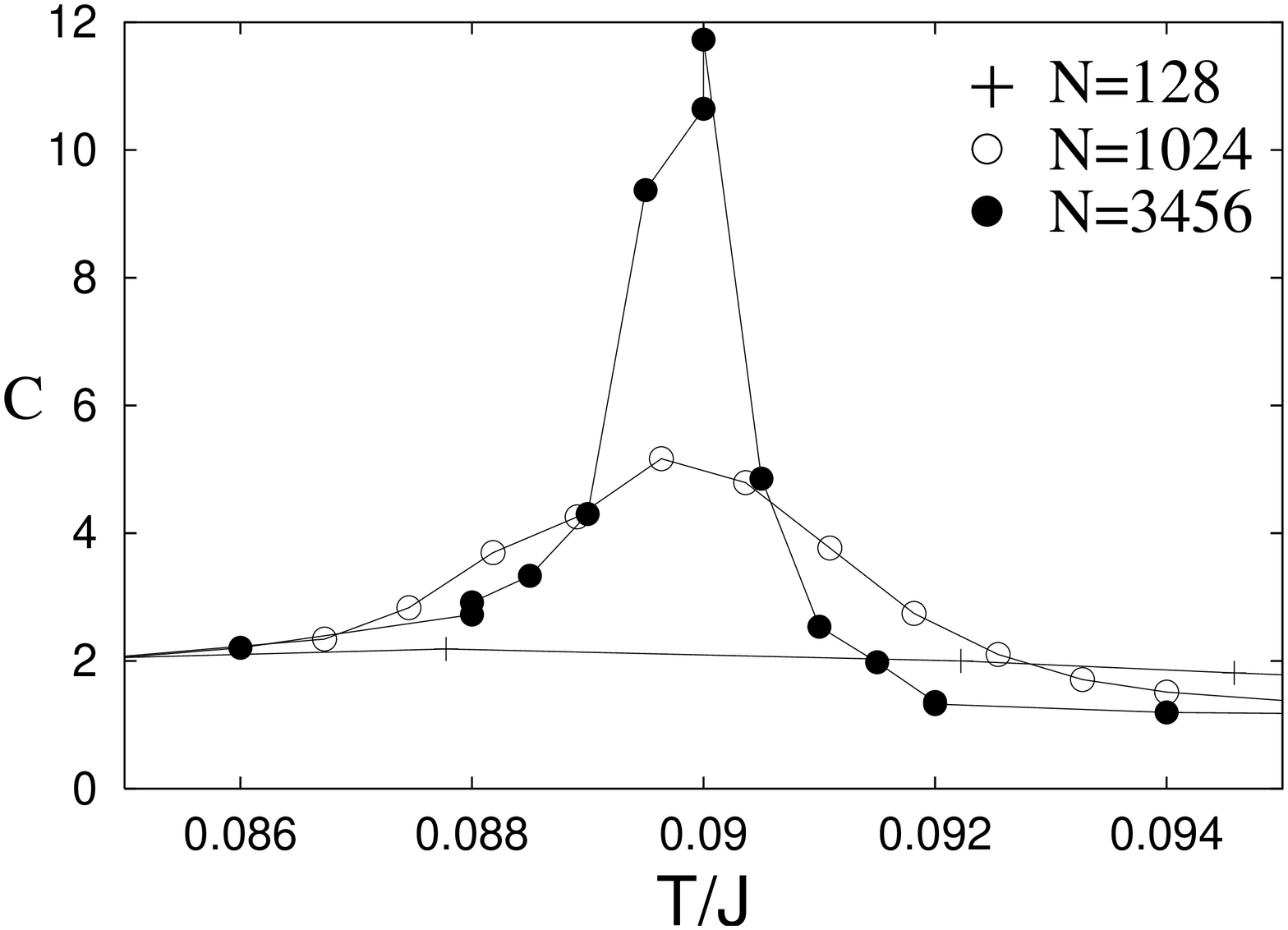}
\caption{\label{fig4}Specific heat for the ``direct'' (up panel) and the ``indirect'' (down panel) cases, obtained in monte carlo simulations for different sizes of clusters.
$\frac{D}{J}=0.1$ and N is the total number of spins.
The critical temperature and more generally the phase transition depends on the type of DMI (note the different scales both in temperature and specific heat ranges between the two cases).
For each point with N=128 and N=1024 (resp. N=3456), $10^6$ (resp. $10^5$) monte carlo steps were disregarded for thermalization and then $10^7$ (resp. $2.10^6$) steps were performed to measure the 
specific heat.
The lines are a guide for the eyes.}
\end{figure}

Monte carlo simulations indicate a phase transition from a high temperature paramagnetic phase to a long range ordered state. 
This is seen in the specific heat (figure \ref{fig4}) and also from snap shots of the magnetic structure at very low temperature. 
The critical temperature was found to be roughly of the order of magnitude of DMI, although it depends quite strongly on the type of DMI (``direct'' or ``indirect'').

Although the DMI are not the more intense interactions in the system, this is not a surprising result for the following reasons. 
In the presence of only nearest neighbour antiferromagnetic interactions (which are the dominant interactions), the pyrochlore lattice 
has a macroscopic degeneracy of its ground state on a mean field level \cite{ReimersBerlinsky}. 
This high degeneracy is not lifted by 
thermal \cite{Reimers, MoessnerPRL} fluctuations and the resulting low temperature magnetic structure is a disordered spin-liquid state. 
In the presence of DMI, even at a mean field level this macroscopic degeneracy is completely lifted. 
The DMI are thus the only interaction responsible for the ordering and this leads to the unusual behaviour that $T_{c}\sim D$ even though DMI are an order of magnitude smaller than $J$ (the isotropic exchange).
This conclusion is also valid in the case of ``indirect'' DMI, although in this case there is a more subtle entropic effect at low temperature as explained in section \ref{indirect}.
Furthermore, in both cases the magnetic structures minimize the isotropic exchange interactions, so there is no competition between DMI and isotropic exchange. 
It is thus expected that $T_c$ is mainly determined by DMI.

\subsubsection{Magnetic structure}

\begin{figure}
\includegraphics[width=7cm]{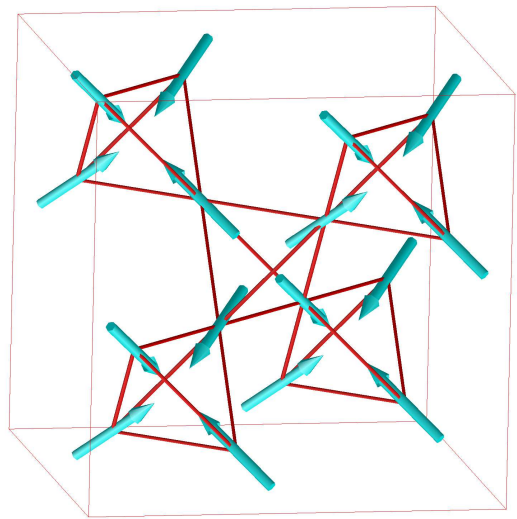}
\caption{\label{fig6}Magnetic structure of the ground state of the pyrochlore lattice in the 
presence of ``direct'' DMI. 
The two structures ``all-in'' and ``all-out'' alternate from one tetrahedron to its neighbours.
}
\end{figure}
In the case of direct DMI (see section \ref{IntroDMI} for a definition of ``direct''), the low temperature magnetic structure is unique except for the time-reversal degeneracy. 
All the spins point towards the center of the tetrahedron or all of them point in the opposite direction and these 
two structures are usually named ``all-in'' and ``all-out''. 
In the pyrochlore structure, the four directions of the spins correspond to the four [111] 
directions in the underlying cubic (fcc) lattice (see figure \ref{fig6}). 
The time reversal degeneracy is still present in the 
pyrochlore structure, but this is only a global degeneracy : the whole structure is uniquely determined as soon as the structure of 
one tetrahedron is fixed to one of the``all-in'' or ``all-out'' states. 
The resulting structure is represented on figure \ref{fig6}, where the two structures ``all-in'' and 
``all-out'' alternate from one tetrahedron to its neighbours. 

\subsubsection{Pyrochlore compounds}

The magnetic structure obtained in the presence of direct DMI (figure \ref{fig6}) was observed experimentally in the pyrochlore compound FeF$_{3}$ 
\cite{Ferey, ReimersGreedan}. 
In this compound, the Fe$^{3+}$ ions have a half filled electronic d shell and an isotropic charge 
distribution with zero orbital momentum (L=0). 
This is unfavorable to the appearance of single ion anisotropies, and in any case can not reasonably explain the observed phase transition at a temperature of 15~K. 
However, DMI are not the only possible origin for this structure, third neighbour interactions \cite{ReimersGreedan} also lead to this ground state. 
It is hard to discriminate between the two scenarii given the experimental and theoretical available predictions. 
However, if DMI are indeed present, this should show up as an effective anisotropy which is not the case with Heisenberg third neighbour interactions. 
It is interesting to note that the parent compound NH$_{4}$Fe$^{\text{II}}$Fe$^{\text{III}}$F$_{6}$ also orders at low temperature and has a magnetic structure 
similar to the one obtained with \emph{indirect} DMI. 
This is detailed in section \ref{indirect}.

\subsection{\label{indirect}Indirect DMI} %

The consequences of indirect DMI are a bit more subtle to study than the direct case due to the occurence of entropic effects (or ``order by disorder'' \cite{Villain}). 
These are not essential to understand the ordering of the system which can be predicted qualitatively on a mean field level, but they are important to predict the 
exact magnetic structure obtained at low temperature. 

\subsubsection{Magnetic structure from monte carlo simulations}

Monte carlo simulations in the presence of indirect DMI indicate a phase transition to a long range magnetic order at low temperature (see the specific heat on 
figure \ref{fig4}). 
The low temperature magnetic structure is a coplanar structure described by a $\mathbf{q}=\mathbf{0}$ wave vector. 
The magnetic moments lie in either the (xy), (zx) or (yz) plane, which represents a global degeneracy.
The magnetic structure of one tetrahedron for the (xy)-coplanar state is represented on figure \ref{fig7}.
Similar structures are easily deduced for the (zx) and (yz)-coplanar structures.

\subsubsection{Mean-field} %

At the very outset, there is no reason to solve the problem within mean-field approximation at $T=0$ since one expects monte carlo simulations to be much more reliable, especially in the 
treatment of thermal fluctuations.
However, the merit of the following mean-field approach is not to reproduce (among others) the magnetic structures found in monte carlo simulations, but precisely to predict magnetic structures which are 
\emph{not} obtained in the simulations. 
This underlines that the magnetic structures observed in the simulations are the result of an entropic selection by thermal fluctuations (which are absent in the mean field treatment).

The structures found in the monte carlo simulations are $\mathbf{q}=\mathbf{0}$ structures~: the magnetic elementary cell is identical to the crystallographic one (one tetrahedron). 
This justifies the following mean field approximation which assumes a wave vector $\mathbf{q}=\mathbf{0}$ and will give the corresponding magnetic structures which minimize the energy. 
The following mean field approach thus neglects any thermal fluctuation (T=0) and consists of minimizing the energy of one tetrahedron with respect to the spins 
coordinates (the eight angles defining the directions or the 4 spins of one tetrahedron). 

\begin{figure}
\includegraphics[width=85mm]{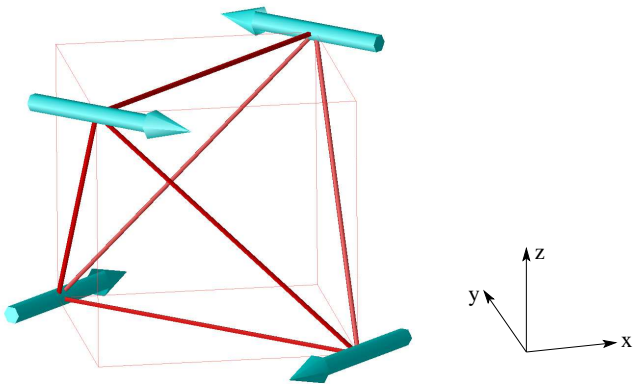}
\caption{\label{fig7}Ground state in the case of indirect DMI.
The ground state for the whole pyrochlore lattice is a $\mathbf{q}=\mathbf{0}$ structure so that only one tetrahedron is represented.
Similar structures in the $zx$ and $yz$ planes are degenerate.
Other non-coplanar states have the same energy but do not participate in the low temperature properties (order by disorder, see text).}
\end{figure}

Doing so, one finds that there is a continuous degeneracy of states which minimize the energy of one tetrahedron. 
These states can be classified in two sets.
The first one is made of the coplanar states obtained in the monte carlo simulations and have a continuous global degree of freedom which is a global rotation in the 
plane (rotation around $z$ on figure \ref{fig7}). 
The second set of lowest energy states contains non coplanar states which can be described starting from a coplanar structure.
Starting from the coplanar structure of figure \ref{fig7}, one can parametrize the non coplanar states as follows~:

\begin{displaymath}
\begin{array}{c}
\mathbf{S}_{1}=\left\{
\begin{array}{l}
\cos\theta\cos\left(\varphi-\frac{\pi}{4}\right)\\
\cos\theta\sin\left(\varphi-\frac{\pi}{4}\right)\\
\sin\left(\theta\right)
\end{array}
\right.
\mathbf{S}_{2}=\left\{
\begin{array}{l}
\cos\theta\cos\left(-\varphi+\frac{\pi}{4}\right)\\
\cos\theta\sin\left(-\varphi+\frac{\pi}{4}\right)\\
-\sin\left(\theta\right)
\end{array}
\right.\\
\mathbf{S}_{3}=\left\{
\begin{array}{l}
\cos\theta\cos\left(-\varphi-\frac{3\pi}{4}\right)\\
\cos\theta\sin\left(-\varphi-\frac{3\pi}{4}\right)\\
-\sin\left(\theta\right)
\end{array}
\right.
\mathbf{S}_{4}=\left\{
\begin{array}{l}
\cos\theta\cos\left(\varphi+\frac{3\pi}{4}\right)\\
\cos\theta\sin\left(\varphi+\frac{3\pi}{4}\right)\\
\sin\left(\theta\right)
\end{array}
\right.
\end{array}
\end{displaymath}

\noindent where the spins are labeled as in figure \ref{fig2} and where $\varphi$ and $\theta$ are \emph{not} independant~: 

\begin{equation}
\theta=\arctan(\sqrt{2}\sin\varphi)
\label{eqn:11h36}
\end{equation}

As soon as (\ref{eqn:11h36}) is true, the corresponding state is one of the degenerate ground states. 
The state represented on figure \ref{fig7} corresponds to $\theta=\varphi=0$. 
In (\ref{eqn:11h36}) $\theta$ is restricted to $[-\frac{\pi}{4}, \frac{\pi}{4}]$, however, starting from a state equivalent to the one of figure \ref{fig7} but where 
the 
spins are coplanar in the (zx) or (yz) plane, one can write down the same kind of parametrization of lowest energy states.
Note that the degeneracy due to the free choice of one of the two angles ($\theta$ or $\varphi$ in equation (\ref{eqn:11h36})) corresponds to a 
global degree of freedom for the pyrochlore lattice, so that the macroscopic 
degeneracy of the pyrochlore antiferromagnet without DMI is lifted and thus a phase transition due to DMI is expected already on a mean field level.

\begin{figure}
\includegraphics[width=85mm]{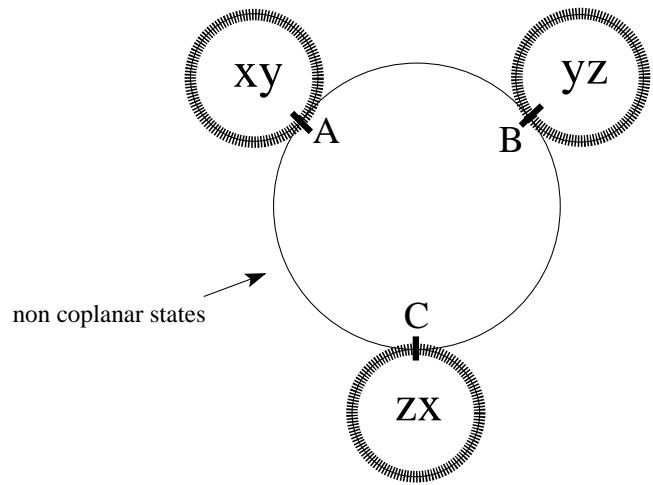}
\caption{\label{fig8}Schematic view of the low lying states in the presence of indirect DMI. (xy), (zx) and (yz) are coplanar states.
Energetically, these states are equivalent to non coplanar states and the magnetic structure can be continuously distorted to explore all these energy minima, as 
represented by the lines of the picture. 
Starting from point A, the thin line can be described by varying $\theta$ and $\varphi$ according 
to Eq.~(\ref{eqn:11h36}) 
(point A : $\theta = \varphi = 0$; 
point B : $\theta = \varphi = - \pi/4$; 
point C : $\theta = \varphi = \pi/4$). 
At finite temperature, the coplanar states are selected by the thermal fluctuations which is represented schematically by broad lines.
}
\end{figure}

The conclusion of the mean field treatment is thus that the coplanar state represented on figure \ref{fig7} is a ground state as well as the equivalent coplanar states in the 
(zx) and (yz) planes, and these three states can be obtained one from another by distorting continuously the magnetic structure while staying at the minimum of the 
energy. 
However, the intermediate states are not coplanar. 
The structure (in phase space) of the lowest energy states is sketched on figure \ref{fig8}. 
As we shall see, the planar and non coplanar states are not equivalent as soon as temperature is not zero~: they all minimize the energy but the thermal fluctuations 
will favour the planar states. 

Finally, the magnetic structure for the direct and indirect cases are very different. 
There is however no reason for them to be related on a frustrated (non-bipartite) lattice. 
For instance the magnetic strucutres for $J>0$ and $J<0$ are very different on the pyrochlore lattice, the ferromagnetic system being magnetically ordered, whereas the antiferromagnetic has a spin-liquid 
ground state.

\subsubsection{Order by disorder}

This discrepancy between the two approaches (mean field and monte carlo) is interpreted as an entropic effect~: the mean field approach neglects the thermal 
fluctuations and identifies the ground states with the minima of the energy. 
However, the fluctuations around the different minima are not equivalent. 
Some of them are entropically favorable, and since at finite temperature the system minimizes its free energy, the 
system will preferably fluctuate around the minima where the entropy is the higher, thus introducing a difference between the degenerate ground states at $T=0$. 
Those states found in monte carlo simulations are a subset of the one found by energy minimization. 
However, this ``order by disorder'' is not responsible for the ordering of the system as 
it only selects coplanar states by removing a global degree of 
freedom (associated with equation (\ref{eqn:11h36})) and fixes the relative orientation of the four N\'eel sublattices, but does not remove a macroscopic number of 
degrees of freedom. 

In order to be more quantitative, a coplanarity parameter (which measures to what extent the system is coplanar) was defined as follows~:

\begin{equation}
\label{13h49}
\mathcal{C} = 1-\frac{3}{N}\min \left(\sum\limits_{\text{spins}}S^{2}_{i,x}, 
\sum\limits_{\text{spins}}S^{2}_{i,y}, \sum\limits_{\text{spins}}S^{2}_{i,z}\right)
\end{equation}

\begin{figure}
\includegraphics[width=85mm]{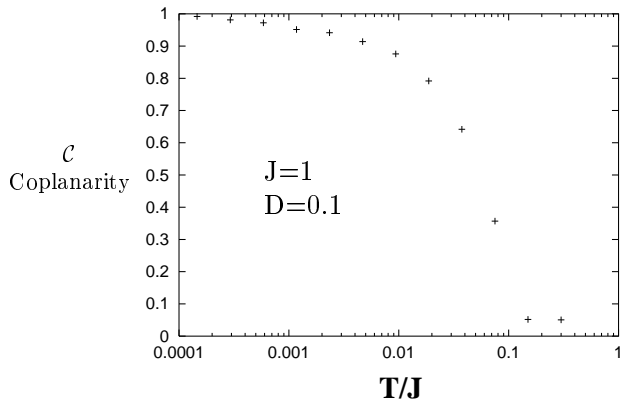}
\caption{\label{fig9}Evolution of the coplanarity parameter $\mathcal{C}$ (see equation (\ref{13h49})) with temperature. 
$\mathcal{C}=1$ corresponds to states where all the spins lie in one of the (xy), (zx) or (yz) planes. 
These coplanar states are selected at low temperature.}
\end{figure}

$\mathcal{C}=1$ in any of the three coplanar states ((xy), (zx) or (yz)), and $\mathcal{C}=0$ in a paramagnetic phase. 
Figure \ref{fig9} represents the evolution of the coplanarity parameter with temperature. 
One clearly sees that the coplanar states are selected. 
From the monte carlo simulations done so far, all the coplanar states (obtained by a rotation around the z axis of figure \ref{fig7}) do no seem to be exactly equivalent, however we could not find a 
clear evidence of a (possible) further order by disorder which would select one particular direction in the selected planes.

\subsubsection{Pyrochlore compounds}

The magnetic structure which appears in the presence of indirect DMI was not observed experimentally as clearly as in the direct case, however, there are two 
compounds whose magnetic structure shares a lot of properties with the one obtained here. 

One of these compounds is very similar to FeF$_{3}$ in which the direct DMI structure is observed. 
This compound is NH$_{4}$Fe$^{\text{II}}$Fe$^{\text{III}}$F$_{6}$ and differs from FeF$_{3}$ by the presence of NH$^{+}_{4}$ which implies both Fe$^{3+}$ and Fe$^{2+}$ magnetic ions with 
the same concentration. 
The latter has one extra electron compared to the half filling case of Fe$^{3+}$.
The observed magnetic structure is similar to the one of figure \ref{fig7}, except that the 90\textsuperscript{o} angle between the spins on the figure is 
$\sim$76\textsuperscript{o} in the compound \cite{FereyBis}. 
Since the Fe$^{2+}$ ion is no more isotropic, it is not surprising that the observed magnetic structure is a little bit different from the perfect DMI structure 
since one expects other sources of anisotropies (like single ion anisotropies) in such a system.
This however suggests that DMI may also be present in NH$_{4}$Fe$^{\text{II}}$Fe$^{\text{III}}$F$_{6}$ but with a change of sign of the DMI compared to FeF$_3$.
 
The other compound whose low temperature long range ordered structure may share similarities with the one found with indirect DMI is the spinel compound 
ZnCr$_{2}$O$_{4}$, although the experimental structure is not uniquely determined \cite{SHL}.
One of the candidates structures has a planar anisotropy in the \{001\} planes and the magnetic structure on one tetrahedron has the 
spins anti-parallel two by two and otherwise perpendicular to each other as shown in 
figure~\ref{fig7}. 
The experimentally observed wave vector is however not $\mathbf{q}=\mathbf{0}$. 
The discrepancy could be due to the structural distortion which occurs in this compound and which lowers the symmetry. 
This is not considered in the present work. 

\subsection{Cu$_{4}$O$_{3}$ (paramelaconite)} %

Cu$_{4}$O$_{3}$ is a mineral compound in which the spin-$\frac{1}{2}$ Cu$^{2+}$ ions form a tetragonally distorted pyrochlore lattice. 
The crystallographic structure is elongated along the (001) direction of the cubic cell. 
On the contrary of what is predicted for the (undistorted) pyrochlore structure with nearest neighbour antiferromagnetic interactions, this compound orders magnetically at low temperature \cite{Pinsard}. 
On-site anisotropies should be small due to the isotropy of the spin-$\frac{1}{2}$ and can not give rise to a critical temperature of $\sim$ 40~K. 
It seemed natural to try and explain the ordering by introducing different coupling constants depending on the direction of the bond, due to the distortion, as well as small interactions between some of the 
second nearest neighbour which should be present in this compound \cite{Pinsard}. 
However, the experimentally observed magnetic structures \cite{Pinsard} could not be reproduced, not even the correct wave vector. 
Experimentally, the magnetic structure is not fully resolved and there are two possible candidates \cite{Pinsard}.
Introducing DMI in accordance with the crystallographic structure, together with the previously tried isotropic exchanges, we were able to reproduce one of the candidates for the magnetic structure. 
More details about this analysis were published elsewhere \cite{Hfm2003}. 

\subsection{\label{QF}Quantum fluctuations} %

All the work presented in this paper treats the magnetic moments as classical 3D vector spins of fixed length. 
This is usually only a good approximation for high spin quantum numbers but is believed to produce at least qualitatively good results for the systems under study 
here, and we now give some clues which support our approach.
First it was shown in a similar study on the kagom\'e lattice that quantum effects are quite small in the presence of DMI interactions \cite{SpinWaves}. 
Second, the pyrochlore lattice is three dimensional and so the quantum effects are expected to be even smaller than in the kagom\'e lattice (2D). 
Third, if the ordering of the paramelaconite (Cu$_{4}$O$_{3}$) is indeed due to DMI as is proposed, then it is an experimental fact that quantum fluctuations 
are not very important in such systems since the magnetic moment of the $S=\frac{1}{2}$ Cu$^{2+}$ ion is still significant \cite{Pinsard}. 
Finally, a related work \cite{Valeri} considered the extreme quantum case of spins $S=\frac{1}{2}$ (using other approximations) and both approaches give similar 
(although not exactly identical) results.

The classical treatment presented here should thus be relevant at least qualitatively even for the extreme quantum case of spins $S=\frac{1}{2}$. 
However, quantum fluctuations were shown to renormalize strongly the critical temperature in a 2D frustrated system of spins $S=\frac{1}{2}$\cite{Luca}, and 
the critical temperature should also be renormalized by quantum fluctuations in the case studied here, although this effect should be weaker because the system is 3D.

\section{Conclusion}

In this paper, the highly frustrated pyrochlore antiferromagnet with additional DMI is studied. 
The relevance of such interactions for this system is shown and the geometry of the \textbf{D} vectors which define them is deduced from symmetry arguments. 
Two kinds of DMI can exist in a pyrochlore lattice and the consequences of both on the magnetic properties are investigated. 
They are found to be in deep contrast with those of the nearest neighbour Heisenberg antiferromagnet on this lattice which has a spin disordered state at low temperature. 
Indeed, as soon as DMI are present, a phase transition to a magnetically long range ordered state will take place at a temperature which is of the order of magnitude 
of the DMI. 
The exact magnetic structures are described and the relevance of this ordering scenario is discussed for compounds in which these structures are experimentally 
observed. 

\begin{acknowledgments}
The authors are very grateful to Seung-Hun Lee and Collin Broholm 
for unpublished results on ZnCr$_2$O$_4$.
It is also a pleasure to acknowledge Loreynne Pinsard-Gaudart, Philippe Monod 
and Juan Rodr\'{\i}guez-Carvajal for helpful discussions of their results on Cu$_4$O$_3$.
\end{acknowledgments}

\end{document}